\documentclass[conference]{IEEEtran}
\IEEEoverridecommandlockouts

\usepackage{cite}
\usepackage{amsmath,amssymb,amsfonts}
\usepackage{algorithm}
\usepackage{algorithmic}
\usepackage{graphicx}
\usepackage{textcomp}
\usepackage{xcolor}
\usepackage{url}
\def\BibTeX{{\rm B\kern-.05em{\sc i\kern-.025em b}\kern-.08em
    T\kern-.1667em\lower.7ex\hbox{E}\kern-.125emX}}
\begin{document}

\title{Enhanced Multi-Class DDoS Attack Identification using a Meta-Learning Ensemble}


\author{Ankith Indra Kumar$^{1}$, Genya Ishigaki$^{2}$,  Amith Kamath Belman$^{3}$
\\Department of Computer Science, College of Science, San Jose State University
\\\textit{\{ankith.indrakumar$^{1}$, genya.ishigaki$^{2}$, amith.kamathbelman$^{3}$\}\texttt{@}sjsu.edu}
}

\maketitle

\begin{abstract}
Distributed Denial of Service (DDoS) attacks continue to pose significant threats to network availability and security. While many detection systems focus on binary classification (attack vs. benign), effective mitigation often requires identifying the specific type of DDoS attack. This paper introduces a robust intrusion detection framework centered around a high-accuracy, multi-class classification model designed to precisely identify various DDoS attack types. We propose an ensemble architecture integrating Long Short-Term Memory (LSTM), K-Nearest Neighbors (KNN), and Random Forest (RF) models, whose outputs are synthesized by a Logistic Regression meta-learner. This approach explicitly addresses the ambiguity often encountered when combining predictions from multiple independent classifiers. Evaluated on the CIC-DDoS2019 dataset, our proposed ensemble meta-learning model achieves 96\% accuracy in the multi-class identification task, significantly outperforming a baseline ``chain model" (combining individual binary classifiers), which reached 92\% accuracy and suffered from high ambiguity. Furthermore, integration and testing within a Software-Defined Networking (SDN) environment using Mininet and the Ryu controller demonstrated the practical applicability of our model, achieving 93\% accuracy in identifying DDoS types in the emulated network traffic. Our work highlights the value of meta-learning ensembles for nuanced DDoS threat identification, paving the way for more adaptive and effective defense mechanisms.
\end{abstract}

\begin{IEEEkeywords}
Distributed Denial of Service,
Ensemble Learning,
Meta-Learner,
Mininet,
Ryu Controller
\end{IEEEkeywords}

\section{Introduction}
Distributed Denial of Service (DDoS) attacks are a persistent and evolving threat to the stability and availability of Internet services. By overwhelming target systems with massive volumes of malicious traffic, often originating from distributed botnets \cite{origpaper1,paper1}, attackers can degrade or completely interrupt critical services. The sophistication of DDoS attacks has grown, ranging from high-volume volumetric attacks saturating bandwidth to protocol attacks exploiting network layer vulnerabilities and stealthy application-layer attacks targeting specific services \cite{paper2}. 
The dynamic and varied nature of these attacks necessitates advanced detection and mitigation strategies that go beyond simple binary identification.

Traditional intrusion detection systems (IDS) and many machine learning (ML) based approaches primarily focus on binary classification – distinguishing malicious traffic from benign traffic \cite{paper3}. While crucial, this binary distinction is often insufficient for implementing the most effective countermeasures. Different DDoS attack types (e.g., TCP SYN flood, UDP flood, ICMP flood) exploit different vulnerabilities and require tailored mitigation responses. Simply knowing an attack is occurring is less actionable than knowing the specific type of attack underway.

Advanced ML offers powerful tools for analyzing complex network traffic patterns and detecting anomalies indicative of DDoS attacks \cite{paper4,paper5}. ML models can learn intricate patterns from vast datasets, adapting to novel attack vectors more effectively than signature-based methods. However, applying ML to multi-class DDoS classification presents unique challenges. A naive approach might involve training separate binary classifiers for each attack type and combining their outputs. We show that this \textit{chaining} strategy often leads to ambiguity, where multiple classifiers trigger simultaneously with similar confidence levels, making precise type identification unreliable.

Software-Defined Networking (SDN) provides a flexible architecture for implementing advanced network security solutions \cite{paper6}. By decoupling the control plane from the data plane, SDN controllers gain a centralized view of the network, enabling dynamic traffic analysis and rapid response. Integrating ML-based detection models with SDN controllers allows for real-time analysis of flow statistics and the automated deployment of mitigation rules based on the identified threat.

This paper addresses the critical need for accurate multi-class DDoS attack identification. We propose a novel ensemble framework that leverages the strengths of diverse base learners, such as Long Short-Term Memory (LSTM), K-Nearest Neighbors (KNN), and Random Forest (RF), and employs a meta-learner (Logistic Regression) to effectively synthesize their predictions and resolve ambiguities inherent in simpler combination methods. 

Our key contributions include
\begin{itemize}
    \item Meta-Learning Ensemble Framework: We design and implement an ensemble model combining LSTM, KNN, and RF, orchestrated by a Logistic Regression meta-learner specifically for multi-class DDoS attack type identification.
    \item Ambiguity Resolution: We demonstrate that the meta-learner significantly reduces classification ambiguity compared to a baseline model that simply chains individual binary classifiers.
    \item Comprehensive Evaluation: We validate our model's performance using both a large-scale, real-world dataset (CIC-DDoS2019) and a realistic SDN emulation environment (Mininet/Ryu). We show that our proposed model outperforms the baseline model in terms of accuracy and F1-score for the multi-class classification task.
\end{itemize}

This paper is an extension of our short paper \cite{our_previous}, which only included the preliminary results with the static CIC-DDoS2019 dataset. Through this extension, we enhanced and implemented our model as an SDN application. The performance of the model was evaluated in the practical SDN emulation environment, which confirms the practicality of the approach.

The rest of the paper is organized as follows: Section \ref{sec:related} discusses related work. Section \ref{sec:method} details our methodology, including the SDN architecture and the ensemble meta-learning model. Section \ref{sec:exp} describes the experimental setup. Section \ref{sec:results} presents and discusses the results. Section \ref{sec:conclusion} concludes the paper and suggests future work.

\section{Related Work \label{sec:related}}

DDoS attack detection strategies have evolved to keep pace with the emerging attack techniques. Early methods relied heavily on signature-based detection, which is effective against known attack patterns but struggles with novel or polymorphic threats. Statistical anomaly detection methods compare network traffic against established baselines of normal behavior, flagging significant deviations \cite{paper7}. While capable of detecting unknown attacks, they can suffer from high false positive rates.

ML has emerged as a promising direction for DDoS detection due to its ability to learn complex patterns and adapt to evolving threats. 
%
\textbf{a)} Supervised Learning: Classifiers like Support Vector Machines (SVM) \cite{paper8,paper9}, KNN \cite{paper9}, Decision Trees, and RF \cite{paper8,paper9} have been widely used, often achieving high accuracy in binary classification tasks (attack vs. benign) on specific datasets. Neural networks, including Multi-Layer Perceptrons (MLPs) and Convolutional Neural Networks (CNNs), have also shown success \cite{paper4}. Recurrent Neural Networks (RNNs), particularly LSTM units, are well-suited for capturing temporal dependencies in network traffic data \cite{paper10}. \textbf{b)} Unsupervised Learning: Clustering algorithms like K-Means and Density-Based Spatial Clustering of Applications with Noise (DBSCAN) can identify anomalous traffic patterns without prior labeling, useful for detecting zero-day attacks \cite{paper11}. \textbf{c)} Ensemble Methods: Simple ensemble techniques of combining predictions from heterogeneous classifiers through averaging or voting have been explored to improve robustness and accuracy over single classifiers \cite{paper12}. However, these often focus on binary classification or use basic combination rules that may not effectively handle multi-class ambiguities.

While many studies focus on binary detection, the need for multi-class classification to identify specific DDoS types (e.g., UDP flood, SYN flood, ICMP flood) is increasingly recognized \cite{paper13,paper14}. Some works have adapted single classifiers (like SVM or decision trees) for multi-class problems. Others have attempted to combine binary classifiers, but often without addressing the challenge of conflicting or ambiguous predictions from different models when faced with data potentially fitting multiple attack profiles partially.

Meta-learning, or stacked generalization, where a meta-model learns to combine the predictions of base models, offers a more sophisticated approach to ensemble construction \cite{paper15,ensemblenew1,ensemblenew2,ensemblenew3}. It has been applied in various domains but less extensively explored for the specific challenge of multi-class DDoS attack type identification, especially concerning the resolution of ambiguity between classifier predictions. While the work in \cite{our_previous} discusses such an ambiguity issue in multi-class classification, they formulate it as a static classification problem for a given dataset.
Our work builds upon existing ML techniques but specifically focuses on constructing a meta-learning ensemble tailored to accurately differentiate between multiple DDoS attack types, addressing the limitations of simpler combination strategies. Furthermore, our model is designed and implemented as an SDN application, which confirms the practicality of the approach. We validate our approach within an SDN context, demonstrating the practical applicability for dynamic network defense.

\section{Methodology \label{sec:method}}
\begin{figure}
    \centering
    \includegraphics[width=0.5\columnwidth]{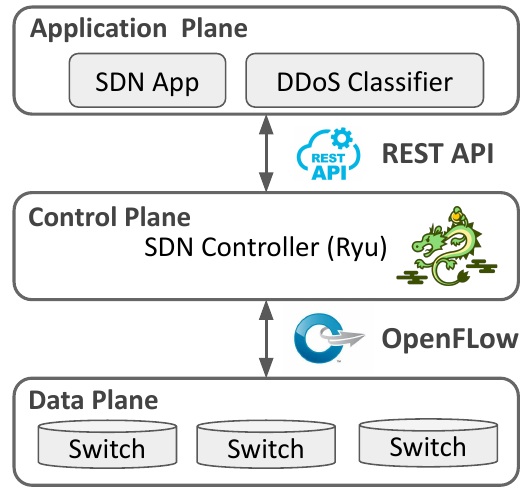}
    \caption{Overview of System Architecture}
    \label{fig:system_arch}
\end{figure}

Our framework integrates an SDN architecture for network monitoring and data collection with an ensemble meta-learning model for multi-class DDoS attack classification. An overview of system architecture is shown in Figure \ref{fig:system_arch}. 

\subsection{SDN-based Data Collection Architecture}
We utilize an SDN architecture to enable centralized network visibility and control, facilitating real-time traffic analysis. 

\begin{itemize}
    \item Components: The architecture employs Mininet \cite{paper16} for network emulation and the Ryu SDN controller \cite{paper17}. Mininet creates a flexible virtual network topology of hosts and OpenFlow-enabled switches. The Ryu controller manages the network devices using the OpenFlow protocol (v1.3).
    \item Data Flow: Network traffic generated within the Mininet environment (both benign and malicious) flows through the virtual switches. The switches, configured with a default table-miss flow entry, forward packets belonging to unknown flows to the Ryu controller.
    \item Feature Extraction: The Ryu controller intercepts these packets and associated flow events (e.g., FlowStatsRequest/Reply). It extracts relevant features from packet headers and flow statistics. We configured Ryu to capture 68 attributes for each flow, including source/destination IP addresses, source/destination ports, protocol types (TCP, UDP, ICMP), packet/byte counts, flow duration, flags, ICMP type/code, etc. This detailed information provides the basis for our ML model's input.
    \item Integration with ML: The extracted flow features are collected and preprocessed before being fed into the ensemble meta-learning model, which resides logically within or is accessible by the control plane. The model's classification output (identifying the specific DDoS type or benign traffic) can then be used to take mitigation actions dynamically programmed by the controller onto the switches.
\end{itemize}

\subsection{ Data Preprocessing}
Raw network flow data requires careful preprocessing to be suitable for ML model training and inference.
\begin{itemize}
    \item Feature Selection: From the initial 68 features captured by Ryu, we selected 21 key attributes deemed most relevant for distinguishing between different DDoS attack types and benign traffic based on prior domain knowledge and preliminary analysis. 

    From the 68 initial features, we first drop any pair whose Pearson $|r| > 0.9$. We then then rank the survivors by Information-Gain and add them one-by-one until the 22nd attribute raised cross-validated F1 by $< 0.5$ \%, leaving 21 dominant features. The final list of features include: Max Packet Length; Fwd Packet Length Max; Total Length of Bwd Packets; Flow Packets/s;  Destination Port; Flow Duration; Flow Bytes/s; Flow IAT Mean; Total Length of Fwd Packets; Bwd Packets/s; Initial Window Bytes (Bwd); Initial Window Bytes (Fwd); Fwd Packet Length Min; Flow IAT Max; Flow IAT Min; Fwd Header Length; Min Segment Size (Fwd); Protocol (Transport); Total Fwd Packets; Source Port;and Total Bwd Packets.

    \item Missing Values: Network data can contain missing values. We employed imputation techniques: mean imputation for numerical features (e.g., packet counts) and mode imputation for categorical features (e.g., protocol type).
    \item Standardization: Numerical features often have vastly different scales (e.g., byte counts vs. flags). We applied standardization 
    to ensure that features with larger values do not disproportionately influence model training, particularly for distance-based algorithms like KNN.
    \item Imbalanced Data: DDoS datasets are typically imbalanced, with benign traffic often vastly outnumbering specific attack types. During the training phase,
    we employed the Synthetic Minority Over-sampling Technique (SMOTE) \cite{paper18} to balance the class distribution. 
\end{itemize}

\subsection{Ensemble Model with Meta-Learner}
Our core contribution is an ensemble model utilizing stacked generalization (meta-learning) to perform multi-class classification of DDoS attacks. The architecture consists of base models and a meta-learner.
\subsubsection{Base Models}
We selected three diverse base models, each capturing different aspects of the data: 

\textbf{LSTM} excels at modeling sequential data and capturing temporal dependencies \cite{paper10}. Network traffic flows have inherent temporal characteristics, and LSTM can learn patterns over time that might distinguish certain attacks or normal behavior. It processes input sequences ($x_t$) that occurs at time $t$, considering previous hidden states ($h_{t-1}$) and cell states ($c_{t-1}$). We use a standard LSTM formulation and optimize it using sparse categorical cross-entropy loss, suitable for multi-class tasks:
$L = -\sum_{i=1}^{N} \sum_{j=1}^{M} y_{ij} \log(\hat{y}_{ij}),$
where \( N \) is the number of samples, \( M \) is the number of classes, \( y_{ij} \) is the binary indicator if class label \( j \) is the correct classification for sample \( i \), and \( \hat{y}_{ij} \) is the model probability prediction for sample \( i \) and class \( j \).

\textbf{KNN} is a non-parametric, instance-based learning algorithm \cite{paper9}. It classifies a new data point based on the majority class among its $k$ nearest neighbors in the feature space, typically determined by a distance metric. KNN is inherently multi-class capable and effective at capturing local patterns in the data. The prediction for a point $x$ is based on a set $N_k(x)$ of its $k$ nearest neighbors:
$  \text{KNN}(x) = \dfrac{1}{k}\sum_{x_i \in N_k(x)} y_i.$

\textbf{RF} is an ensemble method itself, consisting of multiple decision trees trained on different subsets of the data and features \cite{paper8}. It aggregates the predictions of individual trees (typically by majority vote or averaging probabilities) to produce a final classification. RF is robust to overfitting, handles high-dimensional data well, and provides feature importance measures. It naturally extends to multi-class problems. The prediction is an aggregation of individual tree ($T_b$) predictions:
    $\text{RF} = \dfrac{1}{B}\sum_{b=1}^{B} T_b(x;\theta_b),$
where \( B \) is the number of trees, \( T_b \) is the \( b^{th} \) tree, and \( \theta_b \) are the independent identically distributed random vectors. This combination of base models was specifically chosen as LSTM captures temporal dynamics, KNN captures local instance-based similarities, and RF captures complex feature interactions through its tree structures. 

\subsubsection{Meta-Learner}
The predictions from the three base models (LSTM, KNN, and RF) serve as the input features for the meta-learner. We employ Multinomial Logistic Regression (often called Softmax Regression) as our meta-learner.

The meta-learner's role is to learn the optimal way to combine the predictions of the base models. It assesses the confidence and patterns in the predictions from LSTM, KNN, and RF to make a final multi-class classification. Logistic Regression models the probability of a sample belonging to each class $j$ given the input features $X$, which are the predictions from the base models. It uses the softmax function to ensure probabilities sum to 1 across all $K$ classes:   
\begin{equation}
  P(y=j|X) = \frac{e^{X^T \beta_j}}{\sum_{k=1}^{K}e^{X^T \beta_k}},
\end{equation}
where \( P(y=j|X) \) is the probability of class \( j \) given the input \( X \), and \( \beta_j \) is the coefficient vector for class \( j \).
By learning how to weigh and combine the base model outputs, the meta-learner can resolve conflicts or ambiguities where base models might disagree or show similar confidence for different classes, leading to improved overall multi-class accuracy compared to simple averaging or voting.

\begin{algorithm}[t]
\caption{Ensemble Meta-Learning for DDoS Classification in SDN}
\label{alg:ensemble_meta_ddos}
\begin{algorithmic}[1]
\REQUIRE Ryu SDN Controller instance, established connection to Mininet switches, and training flow data (benign and various DDoS types)
\ENSURE A trained meta-learner that produces multi-class predictions (e.g., Benign, DDoS Type 1, Type 2, \dots) and optionally triggers mitigation actions
\STATE \textbf{Initialize:} Ryu SDN Controller
\STATE \textbf{Establish:} Connection to Mininet switches
\STATE \textbf{Training Phase (Offline):}
\STATE \quad Collect flow data via Ryu/Mininet
\STATE \quad Preprocess data 
\STATE \quad Train Base Models on preprocessed data
\STATE \quad Generate predictions from base models
\STATE \quad Train Meta-Learner
\STATE \textbf{Inference Phase (Online):}
\FOR {each incoming flow/packet forwarded to Controller}
    \STATE Extract flow features using Ryu
    \STATE Preprocess extracted features
    \STATE Get predictions from trained Base Models
    \STATE Feed predictions into the trained Meta-Learner
    \STATE Obtain final multi-class prediction from Meta-Learner
    \STATE \textbf{(Optional)} Trigger mitigation based on predicted class
\ENDFOR
\end{algorithmic}
\end{algorithm}

\section{Experimental Setup \label{sec:exp}}
We conducted two sets of experiments to evaluate our proposed ensemble meta-learning model: first using a public benchmark dataset and second using data generated within our SDN emulation environment.

\subsection{Evaluation on CIC-DDoS2019 Dataset}

\subsubsection{Dataset} We used the CIC-DDoS2019 dataset \cite{paper19}, a comprehensive dataset containing benign traffic and various modern DDoS attack types generated using realistic network configurations. It includes detailed flow features extracted using CICFlowMeter-V3.
\subsubsection{Data sampling and preprocessing } From this large dataset, we created a balanced subset for training and testing, focusing on common attack types relevant to our study. We randomly selected approximately 48,000 samples, evenly distributed across Benign traffic and three distinct DDoS attack categories: TCP-SYN Flood, UDP Flood, and ICMP Flood. Preprocessing steps described in the previous section, including feature selection, imputation, normalization, and SMOTE for training data were performed.

\subsection{Evaluation via SDN Emulation (Mininet/Ryu)}

\subsubsection{Network Topology} We constructed a custom network topology using Mininet. The topology consists of 18 host nodes (h1 to h18, each with IP 10.0.0.x/24 and limited CPU) and 6 OpenFlow v1.3 switches (s1 to s6). Hosts were connected directly to individual switches, and switches were connected linearly (s1-s2-...-s6). Note that the topology has a minimal impact on the properties of generated traffic, as all flows are monitored at an attack target.
\subsubsection{Traffic Generation} We used scripting tools (e.g., \textit{hping3}, a custom Python script) running on the Mininet hosts to generate both benign traffic (e.g., HTTP requests, normal pings) and malicious DDoS traffic simulating TCP SYN floods, UDP floods, and ICMP floods directed at specific target hosts within the emulated network.

\subsubsection{Data Acquisition} The Ryu controller, managing the switches, captured flow statistics for the generated traffic using OpenFlow messages, extracting the features. Then, the preprocessing steps described in the previous section are applied, excluding SMOTE during testing/inference.

\subsection{Chain Model}
To demonstrate the advantage of our meta-learning approach, we created a baseline model termed the \textit{Chain Model}. This model represents a more naive strategy for multi-class classification using pre-existing binary classifiers.

\subsubsection{Architecture} The Chain Model utilizes three separate binary classifiers, each highly effective for a specific attack type based on related work \cite{paper8,paper9} and our preliminary tests: SVM for TCP-SYN, RF for UDP Flood, and KNN for ICMP (Ping) attacks. Each classifier is trained to distinguish its target attack type from benign traffic.
\subsubsection{Classification Logic} For a given input sample, predictions (probabilities) are obtained from all three binary classifiers. The model assigns the class corresponding to the classifier with the highest prediction probability.
\subsubsection{Ambiguity Handling} A critical aspect of the Chain Model is its handling of ambiguity. If the difference between the highest and second-highest prediction probabilities falls below a predefined threshold (e.g., 0.1), the sample is classified as \textit{Ambiguous}. We choose the threshold of 0.1 to tune the model on the tighter side since a larger value encourages the chain model to use the ambiguous label more. This indicates that multiple classifiers detected potential attacks with similar confidence, making reliable type identification difficult. This explicit Ambiguous class highlights the limitations of simple chaining.

\subsection{Evaluation Metrics}
\label{subsec:setup_metrics}

We evaluated the performance of both the proposed Ensemble Meta-Learner model and the baseline Chain Model using standard classification metrics:
\begin{itemize}
    \item Accuracy: Overall percentage of correctly classified samples.
    \item Precision: Ability of the classifier not to label a negative sample as positive ($TP / (TP + FP)$).
    \item Recall (Sensitivity): Ability of the classifier to find all positive samples ($TP / (TP + FN)$).
    \item F1-Score: Weighted average of Precision and Recall ($2 \times (Precision \times Recall) / (Precision + Recall)$).
\end{itemize}

\subsection{System Specifications}
\label{subsec:setup_specs}

Experiments were conducted on virtual machines with the following specifications:
\begin{itemize}
    \item Control Plane (Ryu) \& ML Processing: 8-core CPU, 16GB RAM, Ubuntu 20.04 LTS, Python 3.8, Ryu 4.34.
    \item Data Plane (Mininet): 8-core CPU, 16GB RAM, Mininet 2.3.0d6.
\end{itemize}

\section{Results and Discussion}
\label{sec:results}

We present the evaluation results for both the CIC-DDoS2019 dataset and the Mininet/Ryu emulated environment, comparing our proposed Ensemble Meta-Learner model against the baseline Chain Model based on their classification reports and overall performance metrics.

\subsection{Performance on CIC-DDoS2019 Dataset}
\label{subsec:results_cic}

\subsubsection{Chain Model Results}
\label{ssubsec:results_chain_cic}

The performance of the Chain Model on the CIC-DDoS2019 test dataset is summarized in the classification report shown in Table \ref{tab:report_chain_cic}.

\begin{table}[htbp]
\centering
\caption{Classification Report (CIC-DDoS2019) of the Chain Model}
\label{tab:report_chain_cic}
\begin{tabular}{@{}lrrrr@{}}
\hline
Class & Precision & Recall & F1-Score & Support \\
\hline
TCP-SYN Attack     & 0.957 & 0.940 & 0.948 & 3375 \\
UDP Flood Attack   & 0.957 & 0.940 & 0.948 & 3375 \\
ICMP (Ping) Attack & 0.952 & 0.940 & 0.946 & 3375 \\
Benign             & 0.955 & 0.941 & 0.948 & 3372 \\
Ambiguous          & 0.00  & 0.00  & 0.00  & 823  \\ 
\hline
Average & 0.953 & 0.940 & 0.948 & 13498 \\
\hline
\end{tabular}
\end{table}

The Chain Model achieves an overall accuracy of approximately 92\% (if ambiguous samples are counted as incorrect). Precision, Recall, and F1-Scores for the defined attack types and benign class hover around 0.94-0.95. However, the significant support count for the Ambiguous class (823 samples) clearly indicates that the model frequently failed to make a definitive classification, highlighting a major limitation in its practical effectiveness.

\subsubsection{Ensemble Meta-Learner Model Results}
\label{ssubsec:results_ensemble_cic}

The proposed Ensemble Meta-Learner model demonstrates superior performance, effectively resolving ambiguity. Its classification report is shown in Table \ref{tab:report_ensemble_cic}.

\begin{table}[htbp]
\centering
\caption{Classification Report (CIC-DDoS2019) of the Ensemble Model with Meta-Learner}
\label{tab:report_ensemble_cic}
\begin{tabular}{@{}lrrrr@{}}
\hline
Class & Precision & Recall & F1-Score & Support \\
\hline
TCP-SYN Attack     & 0.967 & 0.975 & 0.971 & 3231 \\
UDP Flood Attack   & 0.967 & 0.976 & 0.972 & 3338 \\
ICMP (Ping) Attack & 0.969 & 0.968 & 0.969 & 3361 \\
Benign             & 0.973 & 0.969 & 0.971 & 3568 \\
\hline
Average      & 0.969 & 0.973 & 0.971 & 13498 \\
\hline
\end{tabular}
\end{table}

The Ensemble Meta-Learner model achieves an overall accuracy of 96-97\% (based on Table \ref{tab:report_ensemble_cic}). Precision, Recall, and F1-Scores for all classes are consistently high, generally ranging from 0.96 to 0.97. Crucially, there is no "Ambiguous" class, indicating the meta-learner successfully provided a definitive classification for all samples.

\subsubsection{Comparison}
\label{ssubsec:results_compare_cic}

Table \ref{tab:compare_cic} summarizes the comparison on the CIC-DDoS2019 dataset.

\begin{table}[htbp]
\centering
\caption{Accuracy and F1-Score Comparison (CIC-DDoS2019) }
\label{tab:compare_cic}
\begin{tabular}{@{}lrr@{}}
\hline
Metric     & Chain Model & Ensemble Model \\
\hline
Accuracy   & 92\%        & 96\% \\ 
F1-Score (Avg) & $\sim$0.91-0.94 & $\sim$0.96-0.97 \\
Ambiguity  & High        & Resolved \\
\hline
\end{tabular}
\end{table}

\textbf{Discussion:} The results on the CIC-DDoS2019 dataset clearly demonstrate the superiority of the Ensemble Meta-Learner approach. It not only achieves higher overall accuracy and F1-scores (Table \ref{tab:compare_cic}) but, more importantly, eliminates the ambiguity problem encountered by the Chain Model, as evidenced by the absence of the Ambiguous class with significant support in Table \ref{tab:report_ensemble_cic} compared to that in Table \ref{tab:report_chain_cic}. This indicates that the meta-learner effectively learns to combine the base model predictions for reliable multi-class identification.

\subsection{Performance on Emulated SDN Environment (Mininet/Ryu)}
\label{results_mininet}

\subsubsection{Chain Model Results}
\label{results_chain_mininet}

In the emulated environment, the Chain Model again showed limitations, particularly regarding ambiguity, as detailed in its classification report shown in Table \ref{tab:report_chain_mininet}.

\begin{table}[htbp]
\centering
\caption{Classification Report (Emulated) for Chain Model}
\label{tab:report_chain_mininet}
\begin{tabular}{@{}lrrrr@{}}
\hline
Class & Precision & Recall & F1-Score & Support \\
\hline
TCP-SYN Attack     & 0.89 & 0.90 & 0.89 & 4500 \\
UDP Flood Attack   & 0.87 & 0.88 & 0.87 & 4500 \\
ICMP (Ping) Attack & 0.87 & 0.85 & 0.86 & 4500 \\
Benign             & 0.92 & 0.92 & 0.92 & 5068 \\
Ambiguous          & 0.00 & 0.00 & 0.00 & 1932 \\ 
\hline
Average       & 0.89  & 0.89  & 0.89 & 18568 \\ 
\hline
\end{tabular}
\end{table}

The overall accuracy is around 89\%. Precision, Recall, and F1-scores for the defined classes range from 0.85 to 0.92. The extremely high support count for the Ambiguous class (1932 samples, over 10\% of the non-ambiguous total) severely limits the practical utility of the baseline model in a practical environment.

\subsubsection{Ensemble Meta-Learner Model Results}
\label{ensemble_mininet}

The Ensemble Meta-Learner model maintained strong performance in the emulation, successfully classifying all samples without ambiguity. Its classification report is shown in Table \ref{tab:report_ensemble_mininet}.

\begin{table}[htbp]
\centering
\caption{Classification Report (Emulated) for Ensemble Model with Meta-Learner}
\label{tab:report_ensemble_mininet}
\begin{tabular}{@{}lrrrr@{}}
\hline
Class & Precision & Recall & F1-Score & Support \\
\hline
TCP-SYN Attack     & 0.94 & 0.92 & 0.93 & 4500 \\
UDP Flood Attack   & 0.92 & 0.93 & 0.92 & 4500 \\
ICMP (Ping) Attack & 0.92 & 0.94 & 0.93 & 4500 \\
Benign             & 0.96 & 0.95 & 0.96 & 5068 \\
\hline
Average          &    0.94  &   0.94   & 0.93 & 18568 \\
\hline
\end{tabular}
\end{table}

The model achieves a high accuracy of 93\%. Precision, Recall, and F1-scores are consistently strong, ranging from 0.92 to 0.96 across the classes. Macro and weighted average F1-scores are around 0.93.

\subsubsection{Comparison}
\label{ssubsec:results_compare_mininet}

Table \ref{tab:compare_mininet} summarizes the comparison in the emulated environment.

\begin{table}[htbp]
\centering
\caption{Accuracy and F1-Score Comparison (Emulated)}
\label{tab:compare_mininet}
\begin{tabular}{@{}lrr@{}}
\hline
Metric     & Chain Model & Ensemble Model \\
\hline
Accuracy   & 89\%        & 93\% \\
F1-Score (Avg) & $\sim$0.89      & $\sim$0.93 \\
Ambiguity  & High        & Resolved \\
\hline
\end{tabular}
\end{table}

\textbf{Discussion:} The experiments in the SDN environment confirm the findings from the CIC-DDoS2019 dataset. The Ensemble Meta-Learner model significantly outperforms the Chain Model (Table \ref{tab:compare_mininet}), achieving higher accuracy and F1-scores while successfully resolving the severe ambiguity issue highlighted in Table \ref{tab:report_chain_mininet}. This demonstrates the model's robustness and practical applicability in a dynamic network environment.

The consistent results across both evaluation scenarios strongly support the effectiveness of using a meta-learner to combine diverse base models for accurate and reliable multi-class DDoS attack identification. This approach overcomes the limitations of simpler combination methods, particularly ambiguity, making it a more suitable solution for security systems requiring fine-grained threat classification, as evidenced by the detailed classification reports.

\section{Conclusion and Future Work}
\label{sec:conclusion}

This paper addressed the critical challenge of accurate multi-class DDoS attack identification, moving beyond simple binary detection to enable more effective, tailored mitigation strategies. We proposed a novel intrusion detection framework based on an ensemble of diverse machine learning models (LSTM, KNN, and RF) integrated through a Logistic Regression meta-learner. The framework was designed for and evaluated within an SDN environment using Mininet and Ryu, as well as on the benchmark CIC-DDoS2019 dataset.

Our key findings demonstrate that the proposed Ensemble Meta-Learner model significantly outperforms a baseline \textit{Chain Model}, which naively combines individual binary classifiers. On the CIC-DDoS2019 dataset, our model achieved 96\% accuracy compared to the Chain Model's 92\%, while critically resolving the classification ambiguity that plagued the baseline. In the SDN emulation, our model achieved 93\% accuracy versus the Chain Model's 89\%, again eliminating significant ambiguity. These results highlight the power of meta-learning in effectively synthesizing predictions from heterogeneous base learners for complex classification tasks.

The primary contribution of this work is the development and validation of a robust meta-learning ensemble specifically tailored for multi-class DDoS type identification, demonstrating its superiority over simpler approaches and its applicability within an SDN context. This provides a pathway towards more intelligent and adaptive network defense systems capable of responding precisely to the specific nature of detected threats.

The evaluation included a specific set of DDoS attack types; performance against other, potentially newer or more complex, attack vectors needs further investigation. The reliance on an emulated environment, while controlled, may not perfectly capture all intricacies of real-world network traffic and hardware constraints.
As part of our future work, we plan on expanding DDoS coverage to application-layer and encrypted attacks, and testing real-world performance in network environments. It also involves exploring alternative model architectures, enabling adaptive online learning, and optimizing for resource-constrained deployments.

\end{document}